\documentclass[sn-basic]{sn-jnl}

\usepackage{graphicx}%
\usepackage{multirow}%
\usepackage{amsmath,amssymb,amsfonts}%
\usepackage{amsthm}%
\usepackage{mathrsfs}%
\usepackage[title]{appendix}%
\usepackage{xcolor}%
\usepackage{textcomp}%
\usepackage{manyfoot}%
\usepackage{booktabs}%
\usepackage{algorithm}%
\usepackage{algorithmicx}%
\usepackage{algpseudocode}%
\usepackage{listings}%
\usepackage{subcaption}
\usepackage{enumitem}
\usepackage{amsmath}
\usepackage{multirow}
\usepackage{amssymb}
\usepackage{booktabs}

\theoremstyle{thmstyleone}%
\theoremstyle{thmstyletwo}%

\theoremstyle{thmstylethree}%

\begin{document}


\title[Article Title]{A review on modelling, evaluation, and optimization of cyber-{p}hysical system reliability}

\author[1]{\fnm{Moslem} \sur{Uddin}}\email{moslem.uddin.bd@gmail.com}
\equalcont{These authors contributed equally to this work.}

\author*[2]{\fnm{Huadong} \sur{Mo}}\email{huadong.mo@unsw.edu.au}
\equalcont{These authors contributed equally to this work.}

\author[1,3]{\fnm{Daoyi} \sur{Dong}}\email{daoyidong@gmail.com}
\equalcont{These authors contributed equally to this work.}

\affil[1]{\orgdiv{School of Engineering and Technology}, \orgname{The University of New South Wales}, \orgaddress{\street{Northcott Dr}, \city{Campbell}, \postcode{2610}, \state{ACT}, \country{Australia}}}

\affil*[2]{\orgdiv{School of Systems and Computing Technology}, \orgname{The University of New South Wales}, \orgaddress{\street{Northcott Dr}, \city{Campbell}, \postcode{2610}, \state{ACT}, \country{Australia}}}

\affil[3]{\orgdiv{Australian AI Institute, FEIT}, \orgname{University of Technology Sydney}, \orgaddress{ \postcode{2007}, \state{NSW}, \country{Australia}}}




\abstract{
The aim of this study is to present an overview of current research on modelling, evaluation, and optimization  methods for improving the reliability of Cyber-Physical System (CPS). Three major modelling approaches, namely analytical, simulation, and hybrid models, are  discussed.  Various evaluation techniques, including fault tree analysis, Markov models, and availability measures, are reviewed and compared. Optimization strategies for CPS reliability, including fault tolerance, dynamic reconfiguration, and resource allocation, are also reviewed and briefly discussed. Besides, emerging trends and research opportunities in this field are highlighted and explained. Finally, the possible challenges are outlined and then future research are directed for CPS. This study can provide a systematic and in-dept introduction to CPS for researchers, practitioners, and policymakers.
}




\maketitle

\section{Introduction}\label{sec1}
CPS combines physical components with computational elements, which is becoming increasingly complex and interconnected structures in the recent years. However, it leads to a diverse array of applications across different domains \citep{awotunde2023cyber,8085422}. Therefore, reliability of CPSs are crucial to minimise the potential risks/failures and hence successful operation of CPSs. Recently, several studies have emphasized the significance of CPS in various domains, including smart grid \citep{hasan2023review,habib2023false}, autonomous automobiles \citep{guo2022cyber,zhao2023development}, medical monitoring \citep{ramasamy2022secure,doghri2022cyber}, industrial control systems \citep{zhang2022advancements,saadati2022toward}, and robotics \citep{yun2022immersive,laili2022custom}. Modelling is one of the keys for ensuring reliable operation of CPSs. In literature, various modelling techniques have been explored, including agent-based, actor-oriented, event-oriented, structural/behaviour-oriented, and hybrid approaches \citep{9019636,bemthuis2020using,onaji2019discrete,liu2019reliability,graja2020comprehensive,Nagele2017-ao}.  This literature also explored the challenges associated with CPS modelling. A few of the identified challenges are the need for intuitive abstractions, the adoption of a multi-domain modelling approach, the integration of cyber and physical behaviors, ensuring effective communication and collaboration, representing physical and cyber functionalities, addressing limitations in process modelling capabilities, accounting for timing behavior and constraints, ensuring verification and consistency, and the transformation to mathematical models \citep{graja2020comprehensive,ntalampiras2023few,lee2015past}. 
Nonetheless, there is a noticeable absence of a thorough assessment regarding the effectiveness and constraints of these methods in faithfully illustrating the intricate and dynamic characteristics of CPS. The literature also provides insights into CPS dependability-enhanced optimization methods. However, research on practical implementation and performance and cost tradeoffs is scarce \citep{priyadarshini2022human,cao2021survey,guo2019reliability,fang2017performance}. 
Future research needs to investigate innovative optimization methods that balance reliability, efficiency, and affordability. CPS research provides a foundation for understanding concepts and practical applications, but more comprehensive investigations and in-depth studies are needed.

\subsection{Current Survey}
CPS resilience analysis methods and models were comprehensively reviewed in \citep{cassottana2023resilience}. This study examined a wide variety of resilience analysis models and techniques. 
In \citep{zhang2022advancements} a useful introduction is provided to industrial CPSs. A comprehensive overview of recent advancements in this field, is also presented. This study emphasises the significance and prospective advantages of industrial CPSs. 
The authors of  \citep{alwan2022data} emphasised the importance of data quality management in CPSs and the need for comprehensive solutions to ensure accurate and consistent data. The paper briefly discussed the challenges in assuring the spatial and temporal contextual attributes of sensor node observations in large-scale CPSs. 
The review by \citep{salau2022recent} provided an in-depth investigation of the impact of artificial intelligence (AI) on wireless networking in the context of CPSs and the Internet of Things (IoT). The study briefly covered various research issues. Although it did not fully explore AI for wireless IoT and CPS research directions and trends.
\cite{barivsic2022multi} presented a systematic mapping review of multi-paradigm modelling (MPM) approaches for CPS. The authors aimed to identify and analyze the existing MPM approaches for CPS by investigating the completeness of the approaches, modeled CPS components, employed formalisms, and integration mechanisms. This study briefly discussed the stakeholders involved in CPS modelling. 
A more detailed examination of the roles, responsibilities, and perspectives of different stakeholders would enrich the understanding of the practical implications and challenges of implementing MPM approaches.
A few recent review articles in the existing literature on a similar investigation of CPS are summarized in Table \ref{tab:RecLitRev} to highlight this article’s research scope and significance. Overall, the existing reviews indicate that the information on CPS is scattered throughout the literature. 

\begin{table*}[tp!]
	\centering
	\caption{Examples of existing reviews related to CPS.}
	\label{tab:RecLitRev}
	\begin{tabular}{p{2cm}p{1.5cm}cccp{1.5cm}p{1.5cm}}
		\hline
		Reference &Basic Concept &  Modelling & Evaluation & Challenge & Emerging Trends & Future Direction \\ \hline 		
		\cite{cassottana2023resilience} &  & \checkmark & \checkmark & & &    \\ 
		\cite{zhang2022advancements} &  \checkmark & &  & \checkmark & &   \checkmark  \\ 
		\cite{salau2022recent} &   & &  & \checkmark & &   \checkmark  \\ 
		\cite{lazarova2020reliability} &  \checkmark & &  & \checkmark & &   \checkmark  \\ 
		\cite{yaacoub2020cyber} &   & &  & \checkmark &\checkmark &    \\ 
		\cite{castano2019sensor} &  \checkmark & \checkmark &  & & &   \checkmark  \\ 
		\cite{zhou2019comprehensive} &   & \checkmark &  &\checkmark  & &   \checkmark  \\ 
		\cite{zhou2018review} &   &  &  &\checkmark  & &   \checkmark  \\ 
		This Review & \checkmark & \checkmark  & \checkmark  & \checkmark  & \checkmark  & \checkmark  \\ \hline
	\end{tabular}
\end{table*}

\subsection{Motivation}
In the literature, individual studies have focused on the specific aspects of CPS reliability. However, a comprehensive review consolidating existing knowledge on modelling, evaluation, and optimization is often lacking. The motivation stems from the need to fill this gap in the literature and provide researchers, practitioners, and policymakers with a comprehensive overview of the field.

\subsection{Contribution}
This review paper aims to provide a comprehensive overview of the modelling, evaluation, and optimization of CPS reliability. By exploring existing literature and research contributions, this study  aims to identify key methodologies, techniques, and challenges in this domain, shedding light on the current understanding and potential areas for improvement.
The outcomes of this review paper are expected to benefit researchers, practitioners, and policymakers by providing a comprehensive understanding of the current landscape and the critical aspects involved in modelling, evaluating, and optimizing CPS reliability. Overall, the aim is to contribute to advancing reliable CPS design and operation, ensuring the continued growth and success of these transformative systems.
The novelty of this review lies in the discussion of significant challenges and possible future research directions for CPS.  
In summary, the importance and the originality of this study are as follows:
\begin{itemize}
	\item An overview of the CPS is presented, highlighting its characteristics and challenges related to reliability (Section 2). 
	\item  The modelling of CPS reliability is reviewed by presenting different approaches and considerations for developing precise representations of CPS systems (Section 3). This section also focuses on the evaluation of CPS reliability, exploring metrics, measures, and techniques used to assess the performance and dependability of CPS. 
	\item The optimization of CPS reliability is reviewed by examining the various strategies and trade-offs involved in enhancing system resilience (Section 4). 
	\item The challenges and emerging trends in the field are outlined, along with potential future research directions (Section 5). 
\end{itemize}

\subsection{Paper structure}
\begin{figure}[tp]%
	\centering
	\includegraphics[width=\textwidth]{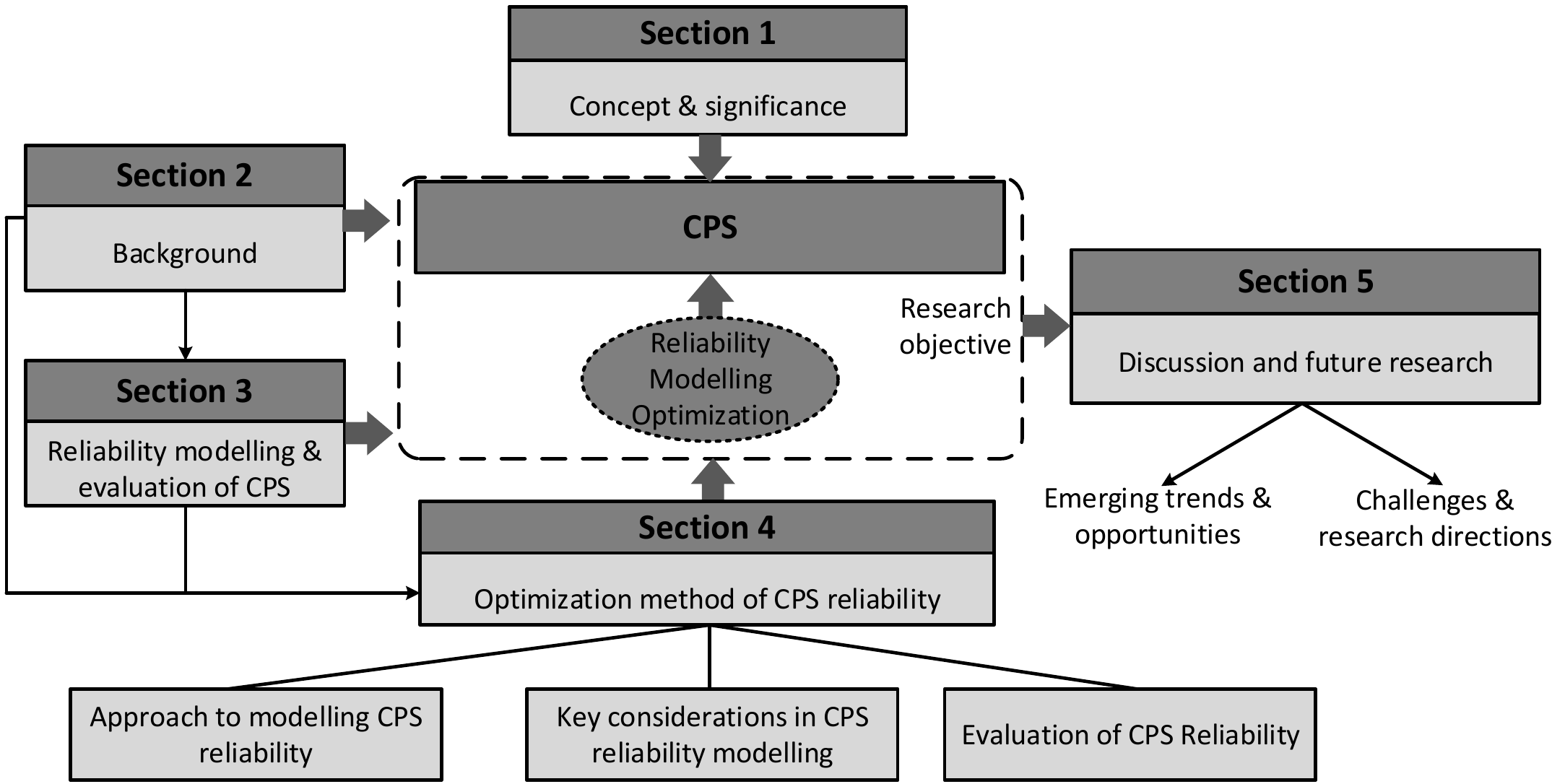}
	\caption{Overall research roadmap for this paper.}\label{fig1a}
\end{figure}

The overall scheme of this paper is shown in Fig. \ref{fig1a}. The remainder of the paper is organized as follows: Section 2 presents an overview of CPS. Section 3 reviews the CPS reliability model and analysis based on various modeling methods and performance measures. In Section 4, the CPS is reviewed based on different optimization methods. The identified challenges, emerging trends, and potential research directions are presented in Section 5. Finally, Section 6 provides the conclusions drawn from the study.
\section{CPS - background}

\subsection{Definition and characteristics of CPS}
A CPS is a system that combines physical and computational elements with communication networks in a tightly integrated manner. An illustrative example of CPS is shown in Fig.\ref{fig1}. CPSs are known for their real-time responsiveness, dynamic behavior, and interaction with the environment \citep{furrer2022cyber}. The key characteristics of CPS are outlined as follows \citep{napoleone2020review}.

\begin{itemize}
	\item \textit{Integration}: 
	CPS integrates physical components with software and communication infrastructure. It results in a seamless integration that unifies the physical and cyber components. There are several physical components including  sensors, actuators, and controllers. 

	\item \textit{Context-dependent behavior}: 
	CPS has context-dependent behavior which enables it to adapt and respond to changes in the environment, user input, and system conditions. This feature makes it effective in dynamic and unpredictable situations \citep{shanaa2017case}.

	\item \textit{Real-time processing}: Real-time processing is frequently required for  CPS  to respond promptly to events or data. Therefore, real-time capabilities are crucial in safety-critical applications where immediate actions are required to prevent accidents or malfunctions.

	\item \textit{Interconnectedness}: CPS components are connected through communication networks. This connectivity helps them to exchange information, share data, and coordinate actions. This feature improves the system performance and enables remote monitoring and management.
	\item \textit{Heterogeneity}: CPS comprises various hardware and software components. Each components is with distinct functionalities, operating systems, and communication protocols. Managing such heterogeneity is challenging in the design and operation of CPS.
\end{itemize}
\subsection{Key components and interactions in CPS}
CPS components can be categorised broadly as physical elements and cyber elements \cite{carreras2020conceptualizing,7815549}:
\begin{figure}[tp]%
	\centering
	\includegraphics[width=\textwidth]{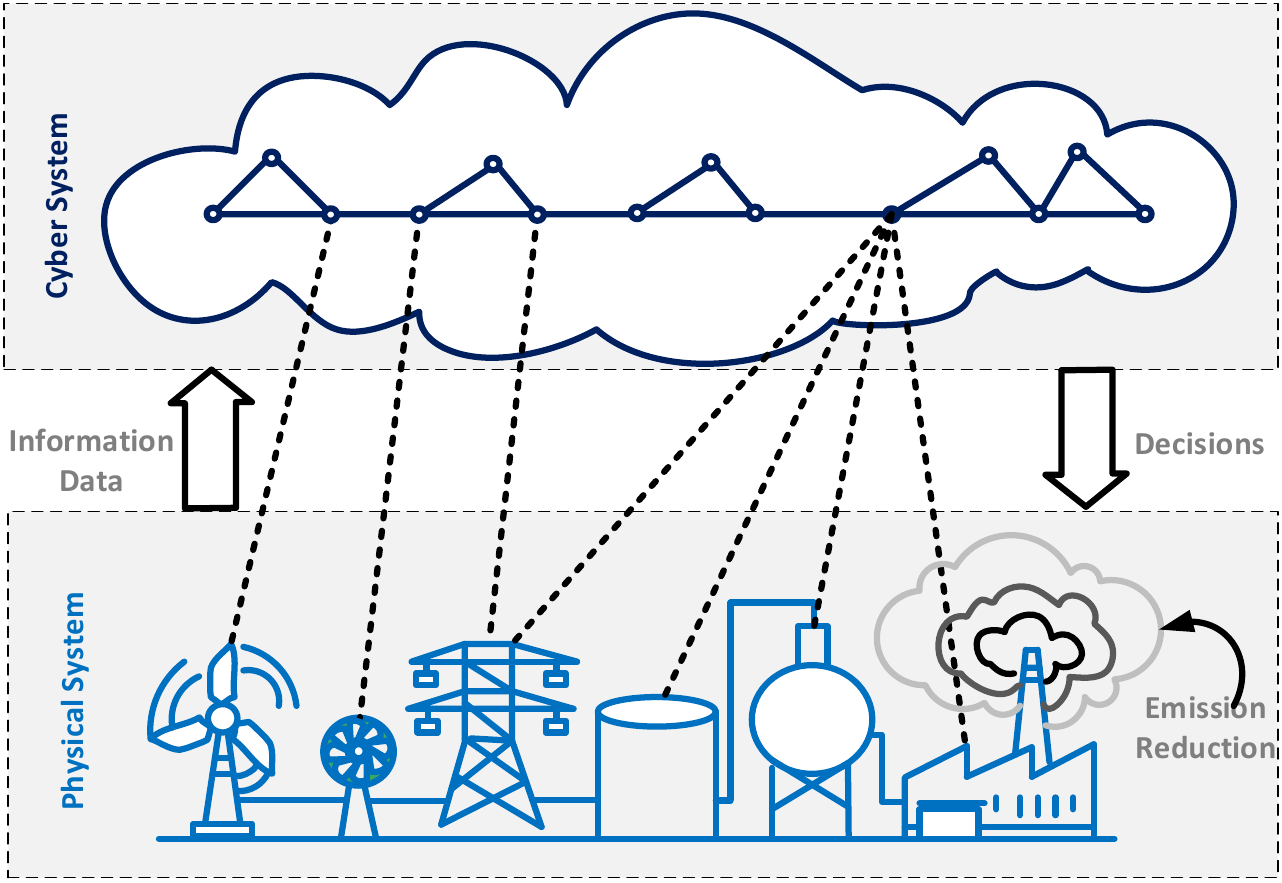}
	\caption{Schematic representation of CPS. The physical system comprises the energy infrastructure, including wind turbines, transmission lines, and industrial facilities. The cyber system processes data collected from the physical system and makes real-time decisions to optimize performance, reduce emissions, and enhance efficiency.}\label{fig1}
\end{figure}
\begin{itemize}
	\item [(i)]\textit{Physical Components}: In a CPS, the physical elements consist of tangible entities that interact with the physical environment. The common physical elements are sensors, actuators, motors, and control systems. Based on the system's computation and control, actuators perform actions. In contract, sensors acquire data from the environment.

	\item [(ii)] \textit{Cyber Components}: In a CPS, various cyber elements are involved. Few common cyber elements are software, communication networks, algorithms, data processing devices, and decision-making systems. These components interpret sensor data, implement control algorithms, and interact with other
	system components.

\end{itemize}

The interactions between the physical and cyber components are continuous and tightly coupled:

\begin{itemize}
	\item \textit{Sensing and perception:} Sensors monitor the physical environment in CPS. They provide important information to the cyber system, which allow cyber system to perceive and interpret the state of the system \citep{wang2022toward}.
	\item \textit{Computation and control:} Components in the cyber realm process sensor data, perform computations, and execute control algorithms. It help to make decisions and generate control signals for the physical elements \citep{7815549,haque2014review}.
	\item \textit{Actuation and effect:} In a cyber system, actuators receive control signals and perform physical actions. Then, it finally affect the surrounding environment \citep{bagula2021cyber,haque2014review}. This concludes the cycle with a sensing phase.
\end{itemize}

\subsection{Applications and domains of CPS}
The CPS field has far-reaching implications across various domains, driving significant advancements in industries such as: 

\begin{itemize}
	\item [--] \textit{Energy}: Integration of renewable energy sources,  smart grids, and energy management systems \citep{hasan2023review,habib2023false}. 
	\item [--] \textit{Smart cities}: Public services optimization, urban infrastructure management,  and environmental monitoring.
	\item [--] \textit{Healthcare}: Monitoring patients remotely, medical devices, and intelligent healthcare facilities \citep{ramasamy2022secure,doghri2022cyber}.
	\item [--] \textit{Transportation}: Technological advances have led to the development of intelligent traffic management, autonomous vehicles, and intelligent transportation systems	 \citep{guo2022cyber,zhao2023development}. 
	\item [--] \textit{Manufacturing}: Smart factories,  industrial automation, and predictive maintenance  \citep{zhang2022advancements,saadati2022toward}.
\end{itemize}

\subsection{Importance of reliability in CPS}
In CPS, safety-critical applications require the highest reliability. To avoid disasters and ensure system performance, it is crucial to operate systems consistently and reliably under different conditions \citep{8666606,fang2017performance}. Reliability directly influences the performance, safety, and overall effectiveness of CPSs. The following factors underscore the significance of reliability in CPS:

\begin{itemize}
	\item [(i)] \textit{Safety and security}:  The reliability of safety-critical CPS is paramount in ensuring the protection of users and the environment, particularly in applications such as autonomous vehicles, medical devices, and industrial control systems \citep{8232537,lyu2019safety}. An unreliable CPS could result in accidents, injuries, or even fatalities. This critical aspect is equally vital for CPS in critical infrastructure to prevent potential disasters and ensure public safety. Power utilities and transportation systems are common examples of such critical infrastructure. 
	
	\item [(ii)] \textit{Continuity of operations}: 
	Critical systems require continuous operation such smart grids, healthcare systems, and industrial automation 	\citep{lee2022sok,alemayehu2017dependability}. These systems face potential downtime or failures that may result in significant economic and societal consequences. Therefore, reliability is crucial to maintain  continuous operation of such systems.
	\item [(iii)] \textit{Efficient resource utilization}: 	A reliable CPS maximizes resource utilization, including energy, materials, and time. For instance, in smart energy systems, a reliable CPS can reduce energy consumption, minimize waste, and effectively manage energy distribution.
	
	\item [(iv)] \textit{Cost-effectiveness}: In CPS, reliability efforts are also crucial for minimizing the financial impact of unexpected downtimes and failures. These events can increase maintenance costs, repair expenses, and productivity losses. Therefore, the objective of reliability efforts is to identify and mitigate vulnerabilities in the system. Additionally, such efforts aim to improve overall performance by reducing the risk of costly disruptions.
	
	\item [(v)] \textit{User trust and acceptance}: For a CPS, user acceptability and confidence are closely tied to its reliability. A greater probability of technology adoption and embrace exists when users possess confidence in the dependability of a CPS \citep{rehman2018effective}. 
	
	\item [(vi)] \textit{Resilience to cyber attacks}:	
	The increasing interconnectedness and vulnerability of CPS to cyber threats underscores the need for enhanced reliability in security. A reliable CPS is better equipped to detect, mitigate, and recover from cyber-attacks. It also preserves system functionality and minimizes potential harm  \citep{colabianchi2021discussing,xu2021bayesian,8289913,xu2022online,mo2023defence}.
	
	\item [(vii)] \textit{Predictive maintenance}: A reliable CPS can support predictive maintenance approaches, allowing operators to anticipate potential equipment failure in advance. In addition, it can take proactive measures to address failures, thereby minimizing unplanned disruptions to operations.
\end{itemize}

To improve the system's reliability and resiliency, this study explores various aspects of modelling, evaluating, and optimizing CPS reliability.

\section{Reliability modelling and evaluation of CPSs}\label{sec2}
\subsection{Approaches to modelling CPS reliability}
Modelling reliability in CPS is essential for understanding system behaviour, predicting failures, and improving system design \citep{mo2021cyber}. Several methods have been used to model the CPS reliability. A comprehensive summary of the pertinent research is presented in Table \ref{tab:LitModelling}. 

\begin{table*}[tp!]
	\centering
	\caption{Related literature analysis on moddelling CPS reliability.}
	\label{tab:LitModelling}
	\begin{tabular}{p{2cm}cp{1.4cm}ccp{6cm}}
		\toprule
		\multirow{2}{*}{References} & \multirow{2}{*}{Year} &  \multicolumn{3}{c}{Modelling Methods} & \multirow{2}{*}{Key Insight}  \\ \cmidrule(lr){3-5} 
		&& Analytical & Simulation & Hybrid &\\ \midrule
		
		\cite{leng2023complex} & 2023 &  &  \checkmark &    & Protecting key nodes can improve the reliability of VPP network effectively.\\
		\cite{jain2021mcs} & 2021 &  &   &   \checkmark & The proposed methodology is investigated on IEEE 24 bus Reliability Test System.\\
		\cite{kumar2021cyber} & 2021 & \checkmark &   &    & The reliability of WSN-based CPS can be quantitatively assessed.\\	
		\cite{peng2020reliability} & 2020 & \checkmark &   &    & A small failure could trigger serious cascading failures within the entire interdependent networks in a CPS architecture.\\	
		\cite{wang2020research} & 2020 &  &   &   \checkmark & Unsupervised learning classification by fast density clustering algorithm can be effectively applied to the importance evaluation of nodes in CPS software system and support the planning of CPS software system.\\

		
		\cite{zhang2019modelling} & 2019 & \checkmark &   &    & The instantaneous availability of the CPS is proved to be fluctuating.\\
		
		\cite{peng2019reliability} & 2019 &  &   &   \checkmark & High degree addition strategy has the best performance in improving interdependent CPS systems after attacked.\\
		\cite{ma2018simulation} & 2018 &  &  \checkmark &    & A fault-tolerant CPS simulation platform was built to validate the method proposed in this paper.\\ 
		
		\cite{lazarova2017reliability} & 2017 &  &   &   \checkmark & Each of the paradigms raises its specific challenges and instills its own types of faults and failures that occur following their specific patterns.\\ \bottomrule
		
	\end{tabular}
\end{table*}

\subsubsection{Analytical models}
Analytical models utilize mathematical equations and probabilistic methodologies to measure CPS's reliability precisely. These models consist of game-theoretic, queuing, fault tree, probabilistic reliability block diagrams, and more \citep{9019636}. They incorporate the dynamics of systems, the potential for malfunctions, and measures of efficiency using mathematical expressions. Analytical models have numerous benefits because of their straightforwardness, efficient computation, and ability to provide analytical perspectives on the dependability of systems.

\subsubsection{Simulation models}  
Simulation models employ computational methods to replicate the temporal dynamics of CPS. Simulation models include, but are not limited to, agent-based models \citep{bemthuis2020using}, discrete event simulation  \citep{onaji2019discrete}, Monte Carlo simulation \citep{liu2019reliability}, and continuous-time simulation \citep{graja2020comprehensive}.  These simulations depict the dynamics and interactions of the components inside a system and their impact on reliability. Simulation models enable the examination of complex situations and the evaluation of system dependability under various operational conditions. In circumstances where analytical solutions are unattainable or when it is critical to represent complex interactions accurately, these methods demonstrate significant benefits.

\subsubsection{Hybrid models} 
Hybrid models integrate components from both analytical and simulation techniques. Analytical techniques are used for specific components or subsystems, whereas simulations are used for more complex or uncertain aspects of the system or the entire system. These models include co-simulation \citep{Nagele2017-ao}, hybrid automata \citep{Rajhans2009-el}, hybrid bond graphs \citep{Simko2014-xc}, etc. Hybrid models effectively harness the advantages inherent in both approaches, facilitating in-depth examination of individual components while simultaneously capturing the holistic behaviour of the entire system.

\subsection{Key considerations in CPS reliability modelling}
To simulate CPS reliability accurately and meaningfully, the following factors need to be considered.
\subsubsection{System architecture and design}
The reliability of CPS significantly depends on the system architecture and design decisions. The arrangement and interactions of components, subsystems, and the overall system architecture should be considered during modelling \citep{lee2008cyber}. To identify potential vulnerabilities and optimize system performance, the selection of architecture and its effect on the reliability of the system must be thoroughly evaluated.

\subsubsection{Component failure models}  
To understand the dependability of CPSs, it is necessary to construct models that simulate component failures. Component failure models comprehensively represent the attributes of failure that affect the overall reliability of a given system, such as repair timeframes, failure rates, and failure modes  \citep{zuniga2020classical,akkaya2016systems}. Several models, including Exponential, Weibull, and Markov models, can be employed to represent component failure probability (FP) \citep{ali2018failure}.

\subsubsection{Interactions and dependencies}  
CPS involves intricate interconnections and interactions between physical and cyber components. It is critical to precisely model these interactions in order to accurately capture the system's reliability behavior. To give a thorough analysis of system dependability, the effect of component breakdowns on the dependability of other components (i.e., reliance) should be taken into consideration.

\subsection{Evaluation of CPS Reliability}

\subsubsection{Metrics for evaluating CPS reliability}

Several metrics and measures can be used to assess the system's dependability and efficacy when evaluating CPS's reliability. A synthesis of relevant literature is presented in Table \ref{tab:LitAssInd}. Meanwhile, Table \ref{tab1} presents a comparative analysis of the key metrics and measures used to evaluate the reliability of CPS. These metrics, including Availability, Mean Time Between Failures (MTBF), and Mean Time to Repair (MTTR), are instrumental in assessing the system performance, resilience, and vulnerability.Table \ref{tab1} provides details of each metric, outlining its purpose and relevant references, and offers a comprehensive basis for evaluating CPS reliability. Some prevalent metrics include the following.

\begin{table*}[htp!]
	\centering
	\caption{Related literature on CPS reliability assessment indices.}
	\label{tab:LitAssInd}
	\begin{tabular}{p{2cm}cp{1.8cm}p{1cm}p{1.2cm}p{1.6cm}p{1.2cm}}
		\toprule
		\multirow{2}{*}{References} & \multirow{2}{*}{Year} & \multicolumn{2}{c}{Evaluation Indices} & \multicolumn{3}{c}{Influencing Factors} \\ \cmidrule(lr){3-4} \cmidrule(lr){5-7}
		&& Standard & New &Physical System Failures& Information Component Failures & Cyber	Attacks \\ \midrule
		
		\cite{zhou2023reliability} & 2023 &  \checkmark &  \checkmark  & \checkmark & \checkmark& \checkmark \\
		\cite{zeng2022analytical} & 2022 &  \checkmark &   & \checkmark & \checkmark& \checkmark \\
		\cite{kumari2022reliability} & 2022 & \checkmark & & \checkmark & &  \\ 
		\cite{chen2021research} & 2021 & \checkmark & & \checkmark & &  \\ 
		\cite{zhang2021application} & 2021 &  \checkmark & & \checkmark & &  \\ 
		\cite{guo2019reliability} & 2019 & & \checkmark & \checkmark &\checkmark &  \\ 
		\cite{chen2018reliability} & 2018 & \checkmark&  & \checkmark &\checkmark & \checkmark \\ 
		\cite{ma2017reliability} & 2017 & &\checkmark  & \checkmark & &  \\ \bottomrule
		
	\end{tabular}
\end{table*}

\begin{itemize}
	\item  \textit{Availability:} The percentage of time when a CPS is operational and able to perform its intended functions is its availability. It considers both scheduled and unscheduled delays to indicate system reliability \citep{Sanislav2019,6578226,Hu2013/11}.
	
	\item \textit{MTBF:} The MTBF of a CPS component is the average time between consecutive failures. It quantifies the reliability of individual components and reveals their failure behaviour \citep{Sanislav2019,6578226,Hu2013/11,10.1115/1.4037228,MOURTZIS2018179}.
	
	\item \textit{MTTR:} The mean time to restore an impaired part or system to an operational state after a failure. It indicates the system's capacity to recover from errors and resume normal operation \citep{Sanislav2019,6578226,Hu2013/11}.
	\item \textit{FP:} The FP refers to the assessment of the probability that a component or the entire system of a CPS will encounter a failure within a specified period \citep{6578226}. A probabilistic viewpoint of system reliability can be offered by this approach, enabling the evaluation of the comprehensive risk associated with the operation of CPS.

	\item \textit{Fault tolerance (FT):} FT quantifies a system's ability to continue operating normally despite component malfunctions or errors \citep{8024524}. It assesses the system's resilience and capacity to degrade gracefully instead of failing completely. A CPS with a high failure tolerance is more reliable and robust.
	
	\item \textit{Mean time to failure (MTTF):} MTTF is the average time a component or system operates before failing. It is commonly used in reliability analysis to estimate components' expected service life or operational lifetime. The longer the operational lifetime and the greater the reliability, the greater the MTTF \citep{Sanislav2019,6578226,Hu2013/11}.
	
	\item	\textit{Failure rate (FR):} Typically expressed as failures per unit of time, FR quantifies the frequency with which failures occur during a specified period \citep{Sanislav2019,6578226,10.1115/1.4037228}. As it provides insight into the severity of failures, it is commonly used to analyse component reliability and predict failure behaviour.
	
	\item	\textit{Reliability block diagram (RBD):} RBD is a graphical depiction of the reliability and stability of system components \citep{8024524}. RBDs aid system reliability analysis and identify critical paths, vulnerable points, and potential constraints. They provide a graphical representation of the system architecture and facilitate the evaluation of the system's reliability and availability.
\end{itemize}

\begin{table}[tp]
	\caption{Comparative table of metrics and measures for evaluating CPS reliability.}\label{tab1}%
	\begin{tabular}{@{}p{1.5cm}p{4cm}p{4cm} p{2.5cm} @{}}
		\toprule
		Metric & Description  & Purpose & Ref \\
		\midrule
		Availability&Measure the percentage of time a CPS is operational&Assess overall system uptime and performance& \citep{Hu2013/11,Sanislav2019}\\
		MTBF &Average time between two consecutive component failures&Compare the reliability of different components&\citep{Sanislav2019,MOURTZIS2018179}\\
		MTTR&Average time required to repair a failed component&Evaluate system recovery time and maintenance efficiency&\citep{Sanislav2019}\\
		FP&Likelihood of a component or system failure&Assess system vulnerability and risk of disruptions&\citep{6578226}\\
		FT&System's ability to function in the presence of failures&Evaluate system resilience and ability to handle errors& \citep{8024524}\\
		MTTF&Average time a component operates before experiencing a failure&Estimate component lifetime and reliability&\citep{Hu2013/11,6578226,Sanislav2019}\\
		FR&Graphical representation of the component reliability and dependencies&Evaluate system architecture and critical paths&\citep{Sanislav2019,10.1115/1.4037228}\\
		\botrule
	\end{tabular}
\end{table}

\subsubsection{Techniques for evaluating CPS reliability}
CPS reliability can be assessed using a range of techniques. Commonly used strategies include the following:

\begin{itemize}
	\item \textit{Fault tree analysis (FTA):} 
	The FTA is a deductive approach that examines potential failure paths within a CPS from a top-down perspective. The methodology utilises a visual depiction of failure events and their interconnectedness to ascertain the combinations of events that lead to failures within the system. The FTA method can clarify the essential paths of failure and assist in understanding how component failures impact the reliability of a system \citep{lazarova2020data,bolbot2020novel}.

	\item \textit{Markov models (MM):} 
	Markov models are stochastic models that represent the dynamic behaviour of a CPS by defining a collection of states and the transitions between them. The researchers quantify the probability of transition between states, considering the dependability of system components and the ramifications of failures. Markov models enable the computation of temporal reliability metrics, such as system availability and component failure probabilities \citep{parvin2013multi,kovtun2022functional,kumar2021cyber}.

	\item \textit{Petri nets (PN):} 
	Petri nets serve as visual depictions of the dynamics and interplay among concurrent processes within a CPS. System reliability can be evaluated by examining the frequency of critical events and their impact on the system's overall performance. Petri nets enable the assessment of reliability metrics and provide insights into the behaviour of CPS \citep{mitchell2013effect,sun2023trustworthiness,li2018reliability,tripathi2021model}.
	
	\item \textit{Failure modes and effects analysis (FMEA):} 
	The FMEA is a proactive approach to identifying and analysing potential failure modes within a given system, product, or process \citep{zuniga2020classical,akula2021risk}. The investigation aims to assess the impact of these failure modes on the overall performance, functionality, and safety of the system. 
	
	\item \textit{Bayesian networks (BN):} 
	The Bayesian network is a graphical model that employs a directed acyclic graph (DAG) to depict a collection of variables and their probabilistic interdependencies. The graph nodes symbolise variables, while the edges represent probabilistic dependencies among these variables. Every vertex in the graph is linked to a conditional probability table (CPT), which specifies the probability distribution of that vertex based on its parent vertices. Bayesian networks can effectively model intricate relationships and inherent uncertainties within components of CPS, including sensors and environmental conditions. Simulations can facilitate the analysis of interrelationships among system variables, evaluation of system reliability, identification of failures, generation of predictions, provision of support for decision-making processes, and mitigation of risks \citep{lyu2020bayesian,almajali2020risk}.
\end{itemize}

\begin{sidewaystable}
	\caption{An overview of different techniques commonly used for evaluating the reliability of CPS}\label{tab3}
	\begin{tabular*}{\textheight}{@{\extracolsep\fill} p{1.5cm}p{2cm}p{3cm}p{3cm}p{2.5cm}p{2cm}}
		\toprule%
		Evaluation Technique & Description	& Key Advantages & Key Limitations & Suitability for CPS Evaluations & Ref  \\
		\midrule
		
		\multirow{2}{*}{FTA} & Analyzes system reliability by constructing a fault tree diagram & (i) Provides a systematic and graphical approach for analyzing failure modes and their causes, and (ii) enables quantitative assessment of system reliability & (i) Large-scale CPS with numerous components and failure modes can make FTA complex, (ii) Constructing accurate fault trees requires expert knowledge, and (iii) Fault trees may not effectively capture dynamic behaviors and dependencies &Well-suited for analyzing and quantifying system reliability and identifying critical failure modes in CPS.& \citep{lazarova2020data,sun2020reliability,niloofar2021fusion}\\
		
		MM   & Models system behavior as a stochastic process based on system states  & (i) Captures dynamic behavior and transitions between states, and (ii) enables analysis of system reliability and availability.  & (i) Assumes memoryless property and exponential distributions, and (ii) limited to systems with discrete states & Suitable for analyzing and modelling CPS reliability, availability, and system dynamics & \citep{parvin2013multi,kovtun2022functional,kumar2021cyber} \\
	
		PN & Graphical modelling technique to represent system behavior and interactions & (i) Captures system dynamics, concurrency, and interactions, and (ii) enables analysis of system reliability and fault propagation & (i) Complexity increases with the size and complexity of the net, and (ii) limited to systems with discrete events & Appropriate for modelling and analyzing CPS with concurrent and asynchronous behaviors & \citep{mitchell2013effect,sun2023trustworthiness,li2018reliability,tripathi2021model} \\
		
		FMEM &Identifies potential failure modes and assesses their effects on system performance& (i) Early identification of failure modes, and (ii) provides insights for risk mitigation and design improvement & (i) Focuses on individual components, and  (ii) may not capture system-level interactions adequately &Highly suitable&\citep{zuniga2020classical,akula2021risk}\\
		
		BN &Probabilistic graphical models that capture dependencies and uncertainty& (i) Captures probabilistic relationships and allows for probabilistic reasoning and inference & (i) Requires prior knowledge and data for accurate modelling, and (ii) complexity increases with the number of variables &Suitable for capturing probabilistic relationships and uncertainties in CPS evaluations&\citep{lyu2020bayesian,almajali2020risk}\\
		\botrule
	\end{tabular*}
\end{sidewaystable}

\subsubsection{Comparison of evaluation techniques and their suitability for CPS}

The choice of evaluation technique is contingent on several variables, including the CPS's complexity, data availability, and the specific reliability concerns being addressed. Each technique has its advantages and disadvantages. Table \ref{tab3} summarises various techniques to aid selection. 

\section{Optimization methods for CPSs reliability}\label{sec3}
\subsection{Optimization objectives for CPS reliability}
Improving the reliability of CPS requires achieving specific objectives that target enhancing both the effectiveness and efficiency of the system. Common CPS reliability optimization objectives are listed as follows \citep{ZHANG20221383,ran2019survey,aslani2022stateoftheart}.

\begin{itemize}
	\item \textit{Maximizing system availability:} 
	The objective of maximising system availability is to minimise system outages and guarantee that the CPS is operational for the desired period. This goal involves minimising planned and unplanned downtime and maximising the system's ability to execute its intended functions \citep{ran2019survey} consistently.
	
	\item \textit{Minimizing FP:}  
	Improving reliability requires minimising the FP of CPS components and the system as a whole. This objective concentrates on decreasing the likelihood of failures occurring within a specified time frame, enhancing the system's ability to operate continuously and without interruption.
	
	\item \textit{Reducing repair and maintenance costs:}  
	In addition to minimising repair and maintenance costs related to CPS failures, balancing maximising reliability and minimising repair and maintenance costs is necessary. This objective seeks to identify strategies for reducing the frequency and duration of maintenance activities without compromising the system's reliability.
\end{itemize}

\subsection{Optimization strategies for CPS reliability} 
The intended objectives of CPS reliability optimization can be achieved through the application of many methodologies and approaches. A few common optimization strategies are discussed as follows.
\subsubsection{Redundancy and FT}  

Redundancy refers to using duplicate or secondary components within a system. This is accomplished by implementing backup systems that provide fault tolerance by allowing the system to continue functioning in the event of individual component failures. Redundancy can be achieved through various means, including hardware redundancy, software redundancy, and data redundancy, each of which contributes to enhancing the reliability and resilience of a system. 
The importance of redundancy has been highlighted in recent research studies \citep{mihalache2019resilience,piardi2020fault}, which underscore redundancy's critical role in ensuring the continuity of operations in various domains.
\subsubsection{Dynamic reconfiguration}  
Dynamic reconfiguration refers to a CPS's capacity to self-adapt and reconfigure in the face of fluctuating operating conditions or malfunctions. CPS enhances system performance and mitigates failures through the dynamic reallocation of resources and modification of system configurations \citep{HEHENBERGER2016273,SANISLAV201667,TOMIYAMA2018161}.

\subsubsection{Resource allocation and scheduling}  
Enhancing resource allocation and scheduling strategies can potentially boost the CPS's reliability. 
This involves efficiently allocating processor power, memory, and communication bandwidth to reduce resource bottlenecks and optimize the system's performance. Effective conflict resolution and resource optimization can be achieved by coordinating activity and task scheduling \citep{5541717,CAPOTA2019204}.

\subsection{Trade-offs between performance metrics in CPS optimization} 
Reliability, system performance, and economic considerations frequently must be compromised to improve CPS reliability. Implementing redundancy mechanisms, allocating additional resources, and engaging in maintenance activities are necessary to increase reliability. These actions can potentially impact the system's performance and incur additional costs. Effectively reconciling and managing these trade-offs is critical for attaining an optimal solution.

\section{Discussion and future research}\label{sec4}
Several research opportunities and unresolved issues have been outlined in the field of CPS reliability. Considering these obstacles and pursuing new research avenues can contribute to the development of CPS reliability. Therefore, this section highlights the challenges in this field and discusses the potential research directions in detail.

\subsection{Emerging technologies and research opportunities in CPS}
The emerging trends and technologies that have the potential to improve CPS reliability are discussed bellow, along with the associated research opportunities.
\begin{itemize}

\item \textit{AI and machine learning (ML):} 
Recently AI and ML techniques are gaining much attention in the filed CPS reliability research \citep{salau2022recent}. 
These techniques can analyse huge data from CPS sensors and components. It supports in identifying patterns, anomalies, and potential failure precursors. This enables the implementation of proactive maintenance strategies, predictive failure detection, and adaptive system behaviour. It enhances the reliability of CPS by allowing early intervention and prompt response to potential failures. Therefore, it will be interesting to explore the application of AI and ML techniques in CPS for predictive maintenance, anomaly detection, optimization, and decision-making processes.
	
\item 	\textit{Edge and fog computing:} 
CPS frequently operates in distributed and decentralized environments. Therefore, it requires real-time processing and decision-making. Emerging edge and fog computing paradigms introduce computation and storage capabilities closer to the network's edge. This proximity reduces latency and allows faster response. Edge and fog computing can facilitate CPS by analyzing data on time, offering local decision-making, and providing resilient operation despite communication disruptions. These capabilities show that edge and fog computing can improve CPS's reliability \citep{de2018application,tran2023fog}. Therefore, future research should explore integrating this technology in CPS to enable decentralized processing, reduce latency, and improve overall system performance and reliability.
	
\item \textit{Blockchain technology:} 
Blockchain technology can improve CPS reliability by providing 
distributed consensus, immutability, and transparency  \citep{aiden2023ai,biswas2023study}. CPS can establish secure and reliable communication and data exchange protocols using this technology. Smart contracts based on blockchain technology in CPS can facilitate auditable and reliable interactions between CPS components. It will 
mitigate cybersecurity risks and thereby enhance system reliability. It is, therefore, essential to explore the use of blockchain and other distributed ledger technologies for securing data transactions, enhancing trust, and improving the integrity of information in CPS applications.
	
\item \textit{Resilient and self-healing systems:} 
Resilience is becoming an important component of CPS reliability. To maintain core functionality, resilient CPS systems are capable of detecting, adapting to, and recovering from failures. CPS is able to withstand failures, recover from them, and continue operating with minimal delay or performance degradation through techniques such as FT, self-healing mechanisms, and dynamic reconfiguration. These approaches centred on resilience improve the overall reliability of CPS systems \citep{balchanos2023towards,loh2023enhancing,yadav2023resilient}. It is essential to explore self-healing mechanisms within the context of CPS, particularly considering adaptive control strategies, autonomous system reconfiguration, and ML-based methods for dynamic adaptation in response to failures.
	
	\item \textit{Digital twins:} 
	Digital twins are virtual copies of physical CPS systems that facilitate real-time monitoring, analysis, and simulation. Creating a digital counterpart of a CPS enables continuous monitoring and evaluation of system behaviour, failure prediction, and performance OP. Digital twins enable proactive maintenance, rapid testing of system modifications, and parameter OP, thereby augmenting CPS reliability \citep{marah2023architecture,de2023exploratory}. Investigating the implementation of digital twins in CPS involves exploring the creation of virtual representations of physical systems for simulation, monitoring, and real-time analysis, with the aim of improving system understanding and reliability.
	
	\item \textit{Human-centric design:} 
	The importance of recognising the role of human operators and users in CPS reliability is growing \citep{khari2023guest}. In CPS design and operation, human-centric design principles seek to understand and account for human factors, including cognition, decision-making, and usability. Taking into account the capabilities, limitations, and behaviour of human operators can increase the reliability and safety of CPS by decreasing the likelihood of human error and enhancing human-system interaction. 
	Undertaking research on human-centric security approaches for CPS, while simultaneously grappling with the challenges posed by cybersecurity measures and their ramifications for end-users, represents a fascinating area of study. It is crucial to explore strategies that can bolster security without compromising the user experience. 
	
	\item \textit{Quantum Computing for CPS:}   Exploring the prospects of employing quantum computing to tackle intricate optimization issues, strengthen cryptographic systems, and bolster the dependability and security of CPS.
	
	\item \textit{Explainable AI for CPS:} Undertaking research to enhance the transparency and interpretability of AI and ML models within CPSs, with a particular focus on addressing the requirement for explainable AI in high-stakes applications where a thorough comprehension of decision-making is essential.
	
\end{itemize}

\subsection{Challenges and future research directions }
Despite the potential benefits, CPS has several challenges and limitations, as explained. The
fundamental challenges of CPS can be classified as follows.
\begin{itemize}
	\item \textit{Integrated modelling and evaluation:} 
	CPS exhibits complex interactions and interdependencies among its physical and cyber constituents, posing challenges in accurately modelling reliability. The development of integrated frameworks that incorporate analytical and simulation-based approaches has potential to address this complexity. These frameworks must capture the intricate relationships among components, subsystems, and the entire system, thus ensuring a thorough and comprehensive analysis of the reliability of CPS.
	
	\item \textit{Multi-layered security and reliability:} 
	The interplay between security and reliability in CPS is a difficult and complex topic. Future research should investigate the incorporation of security and reliability measures to mitigate potential threats and address vulnerabilities. The development of techniques for evaluating the impact of security measures on CPS reliability and the investigation of tradeoffs between security and reliability objectives are crucial for the development of reliable and robust CPS systems.
	
	\item  \textit{Data-driven reliability analysis:} 
	Utilizing the vast amount of data generated by CPS sensors and components poses numerous challenges for enhancing reliability analysis. Exploring advanced data analytics techniques, such as anomaly detection, pattern recognition, and prognostics, can improve the identification of failure patterns, prediction of failures, and optimization of maintenance strategies. Integrating real-time data into reliability models and analysis techniques will allow for more accurate and responsive assessments of reliability.

	\item \textit{Incorporating dynamic and uncertain environments:} 
	CPS functions within the dynamic and unpredictable contexts, characterised by the potential for rapid fluctuations in conditions. The task of modelling and assessing reliability in such circumstances presents significant challenges, primarily stemming from the necessity to incorporate uncertain inputs, fluctuating workloads, and dynamic system configurations. Future research should focus on resilient methodologies that effectively encompass the intricacies and ambiguities of the surroundings to ensure reliable CPS operation.

	\item \textit{Trade-offs between performance metrics:} 
	Optimal CPS reliability frequently necessitates trade-offs with other performance metrics, including cost, energy efficiency, and system responsiveness. The balancing of these competing goals and the optimization of CPS reliability, considering resource limitations and operational requirements, is a challenging task. For reliable CPS design and operation, it is essential to develop optimization techniques that consider these trade-offs and provide applicable solutions.
	
\end{itemize}

\section{Conclusion}
This study presents a comprehensive review of the modelling, evaluation, and optimization of CPS reliability. 
This review classifies modelling approaches into three groups, namely analytical, simulation, and hybrid models. Key aspects of these techniques are also reviewed. In addition, this study covers several evaluation techniques including fault tree analysis, Markov models, and availability measures. Further, optimization strategies are reviewed to provide an insight into these techniques. This review also highlights recent trend in the field of CPS. A number of research issues and challenges have been identified for CPS. 
Future research areas to address the identified issues and challenges have been outlined. The state-of-the-art information provided in this review would draw attention to the investigators, experts, and researchers for CPS.

However,  this study has not thoroughly addressed CPS security issues such as attacks and vulnerabilities. Therefore, it would be interesting to explore security aspects including threat modelling and vulnerability analysis as future extension of this study. 

\bibliography{sn-bibliography}

\end{document}